\documentclass[a4paper,12pt]{article}
\usepackage{amsthm}
\usepackage{amssymb}
\usepackage{amsmath}
\usepackage{amsfonts}

\newtheorem{lemma}{Lemma}[section]
\newtheorem{theorem}{Theorem}[section]
\newtheorem{proposition}{Proposition}[section]
{\tiny {\LARGE }}

\def\b1{\mbox{\boldmath $1$}}

\newenvironment{demo*}{\vspace{3mm}\noindent{\bf Proof.}}{\hfill $\Box$ \vspace{3mm}}

\begin{document}

\baselineskip=22pt
\title{\bf \Large Optimal dividend problems for a jump-diffusion  model
 with capital injections and proportional transaction costs}
 \author{\normalsize $^a${\sc Chuancun Yin } and $^b${\sc  Kam Chuen Yuen }\\
{\normalsize\it $^a$  School of  Statistics, Qufu Normal University,}\\
{\normalsize\it Shandong 273165, P.R.\ China } \\
{\small\it  Corresponding author: E-mail: ccyin@mail.qfnu.edu.cn  } \\
[3mm] {\normalsize\it $^b$ Department of Statistics and Actuarial
Science, The University of Hong Kong,}\\
{\normalsize\it Pokfulam Road, Hong Kong }\\
{\small\it E-mail:  kcyuen@hku.hk}}  \maketitle
\noindent {\bf Abstract}  \ \ \ In this paper, we study the optimal control problem for a company whose
surplus process evolves as an upward jump diffusion with random return on investment. Three types of practical optimization problems faced by a  company that can control its liquid reserves by paying dividends and injecting capital.
In the first problem, we consider the classical dividend problem without capital injections.
The second problem aims at maximizing the expected discounted dividend payments
minus the expected discounted costs of capital injections over strategies with positive surplus at all times. The third problem has the same objective as the second one, but without the constraints on capital injections. Under the assumption of proportional transaction costs, we identify the value function and the optimal strategies for any distribution of gains.

\noindent {\it Key words and phrases}. {Barrier strategy, dual model, HJB equation,   jump-diffusion,  optimal
dividend strategy,      stochastic control.

}


\normalsize

\baselineskip=22pt

\section{\normalsize INTRODUCTION}\label{intro}
For the optimal dividend problem, one may adopt the objective of maximizing the expectation of the
discounted dividends until possible ruin. This problem was first addressed by De Finetti [16]
who considered a discrete time risk model with step sizes $\pm 1$ and showed that the optimal dividend
strategy is a barrier strategy.  Miyasawa [21] generalized the model to the case that periodic
gains of a company can take on values $-1,0,1,2,3,\cdots$, and showed that the optimal
dividend strategy of the generalized model is a barrier one.  Subsequently, the problem of finding the
optimal dividend strategy has attracted great attention in the literature of insurance mathematics.
 For nice surveys on this topic, we refer the reader to Avanzi [3] and Schmidli [22].
  Besides insurance risk models, the optimal dividend problem in the so-called dual model has also been
 studied extensively in recent years.  Among others, Avanzi et al. [6] discussed how the expectation
     of the discounted dividends until ruin can be calculated for the dual model when the gain amounts follow an
     exponential distribution or a mixture of exponential distributions, and showed how the exact value of the
     optimal dividend barrier can be determined; and Avanzi and Gerber [5] examined the same problem for
    the dual model that is perturbed by diffusion, and showed that the optimal dividend strategy in the dual model
    is also a barrier strategy.   To make the problem more interesting, the issue of capital injections has also
    been considered in the study of optimal dividends in the dual model.  Yao et al. [23] studied the optimal
    problem with dividend payments and issuance of equity in the dual model with proportional transaction costs,
    and derived the optimal strategy that maximizes the expected present value of dividend payments minus
    the discounted costs of issuing new equity before ruin.  Yao et al. [24] considered the same problem
    with both fixed and proportional transaction costs.  Dai et al. [14,15] investigated the same problem as in Yao et al. [23] for the dual model with diffusion with bounded gains and exponential gains, respectively.  Avanzi et al. [7] derived an explicit expression for the value function in the dual model with diffusion when the gains distribution in a mixture
    of exponentials in the presence of both dividends and capital injections. Specifically, they showed that barrier
    dividend strategy is optimal, and conjectured that the optimal dividend strategy in the dual model with diffusion
    should be the barrier strategy regardless of the distribution of gains.  Bayraktar et al. [11] examined the
     same cash injection problem, and used the fluctuation theory of spectrally positive L\'evy processes to show
    the optimality of the barrier strategy for all positive L\'evy processes. Bayraktar et al. [12] extended the study to the case with fixed transaction costs. Other related work can be found in Yin and Wen [26], Yin, Wen and Zhao [28], Avanzi et al. [8],  Yao et al. [25] and Zhang [29].

In this paper, we provide a uniform mathematical framework to analyze the optimal control problem with dividends and capital injections in the presence of proportional transaction costs for the dual model  with random return on investment.  The associated value function is defined as the expected present value of dividends minus costs of capital injections until ruin. The rest of the paper is organized as follows.  In Section 2, we give a rigorous mathematical formulation of the problem. Section 3 works on the model without capital injections, while Section 4 deals with the model with capital injections which never goes bankrupt.
Finally, we solve the general stochastic control problem in Section 5.

\section{\normalsize Problem formulation}\label{math}

Assume that the surplus generating process $P_t$ at time
$t$ is given by
\begin{equation}
P_t=x-pt+\sigma_p W_{p,t}+\sum_{i=1}^{N_{t}}X_{i},\ \ t\ge 0,
\end{equation}
 where $x>0$ is the initial assets, $p$ and $\sigma_p$ are positive constants,  $\{W_{p,t}\}_{t\ge 0}$
is a standard Brownian motion independent of the homogeneous compound
Poisson process $\sum_{i=1}^{N_{t}}X_{i}$, and $\{X_{i}\}$ is a sequence of independent and identically distributed random variables having common distribution function $F$ with $F(0)=0$.   Let $\lambda$ be the intensity of the Poisson process $N_{t}$. We assume throughout the paper that
$E[X_{i}]<\infty$ and $\lambda E[X_i]-p>0$.
Here, we consider the return on investment generating process
\begin{equation}
R_t=rt+\sigma_R W_{R,t},\ \ t\ge 0,
\end{equation}
where  $\{ W_{R,t}\}_{t\ge 0}$ is another standard Brownian motion, and $r$ and $\sigma_R$ are positive constants. It is assumed that $W_{p,t}$ and $W_{R,t}$ are correlated in the way that
$$W_{R,t}=\rho W_{p,t} +\sqrt{1-\rho^2}W_{p,t}^0,$$
where $\rho\in [-1,1]$ is constant, and  $W_{p,t}^0$ is a standard Brownian motion independent of $W_{p,t}$.

Define the risk process $U_t$ as the total assets of the company at time $t$, i.e., $U_t$ is the solution to the stochastic
differential equation
  \begin{equation}
  U_t=P_t+\int_0^t
U_{s-}\text{d}R_s,\;\; t\ge 0.\label{math-eq0}
\end{equation}
The solution to (2.3) is given by (see, e.g.  Jaschke [19, Theorem 1])
\begin{equation}
U_t={\cal E}(R)_t\left(x+\int_0^t{\cal E}(R)_{s-}^{-1}\text{d}P_s-\rho\sigma_p\sigma_R \int_0^t{\cal E}(R)_{s-}^{-1}\text{d}s\right),\nonumber
\end{equation}
where
$${\cal E}(R)_t=\exp\{(r-\frac12\sigma_R^2)t+\sigma_R W_{R,t}\}.$$

Using It$\hat{\text{\rm o}}$'s formula for semimartingale, one can show that the infinitesimal generator
$\cal{L}$ of $U=\{U_t,t\ge 0\}$ is given by
 \begin{eqnarray}
  {\cal L} g(y)=(ry-p) g'(y)
  &+& \frac12 \left[(\sigma_p+\rho \sigma_R y)^2+\sigma^2_R (1-\rho^2)y^2\right]g''(y)\nonumber\\
&+&\lambda\int_{0}^{\infty}[g(y+z)-g(y)]F(\text{d}z). \label{math-eq1}
 \end{eqnarray}
The model (2.3) is a natural extension of the dual model in  Avanzi and Gerber [5] and  Avanzi et al. [6].
As was mentioned in Avanzi et al. [6], the dual model is appropriate for companies that have deterministic expenses and occasional gains whose amount and frequency can be modelled by the jump process $\sum_{i=1}^{N_{t}}X_{i}$. For example, for companies such as pharmaceutical or petroleum companies, the jump could be interpreted as the net present value of future gains from an invention or discovery. Another example is  the venture capital investments or research and development investments. Venture capital funds screen out start-up companies and select some companies to invest in. When there is a technological breakthrough, the jump is generated. More examples can be found in Bayraktar and Egami [10] and Avanzi and Gerber [5].

 In this paper, we  denote by $L_t$ the cumulative amount of dividends paid up to  time $t$ with $L_{0-}=0$, and by $G_t$ the total
amount of capital injections up to  time  $t$ with
$G_{0-}=0$. A dividend control strategy $\xi$ is described by the stochastic
process $\xi=(L_t, G_t )$. A strategy  is called admissible if  both
$L$ and $G$ are non-decreasing $\{\cal{F}$$_t\}$-adapted processes,  and
their sample paths are right-continuous with  left limits.  We denote by $\Xi$  the set of all
admissible dividend policies. The risk process with
initial capital $x\ge 0$ and controlled by a strategy $\xi$ is given
by $U^{\xi}=\{U_t^{\xi}, t\ge 0\}$, where   $U_t^{\xi}$ is  the solution to the stochastic differential equation
\begin{equation}
\text{d}U^{\xi}_t=\text{d}P_t+ U^{\xi}_{t-}\text{d}R_t-\text{d}L_t^{\xi}+\text{d}G_t^{\xi},\;\; t\ge 0.\nonumber
\end{equation}
Moreover,
$L_{t}^{ \xi}-L_{t-}^{ \xi}\le U_{t-}^{ \xi}$ for all $t$. In
  words,  the amount of dividends is smaller than the size
of the available capitals. Let $\tau^{\xi}=\inf\{t\ge 0:
U^{\xi}_t=0\}$ be the ruin time. Then, the associated performance function is given by
\begin{equation}
V(x;\xi)=E_x\left(\alpha\int_{0-}^{\tau^{ \xi}-}\text{e}^{-\delta
t}dL_t^{ \xi}-\beta \int_{0-}^{\tau^{ \xi}-}\text{e}^{-\delta t}dG_t^{
\xi}\right),\label{math-eq4}
\end{equation}
 where $\delta>0$ is the discounted rate, $1-\alpha$ ($0<\alpha\le 1$)  is the rate of proportional costs on dividend transactions,
 $1\le \beta<\infty$ is the rate of proportional transaction costs of capital injections. The notation $E_x$
 represents the expectation conditioned on $U^{\xi}_0=x$ and the
integral is understood pathwise in a Lebesgue-Stieltjes sense. Our
aim is to find the value function
\begin{equation}
V_*(x)=\sup_{ \xi\in\Xi}V(x; \xi),\label{math-eq5}
\end{equation}
and the optimal policy $ \xi^*\in\Xi$ such that $V(x;\xi^*)=V_*(x)$
for all $x\ge 0$.

The study of optimal dividends has been around many years. The commonly-used approach to solving these
optimal control problems is to proceed by guessing a candidate optimal solution, constructing the corresponding value function, and subsequently verifying its optimality through a verification result.  For the model of study, i.e., an upward jump-diffusion process with random return on investment, the optimal control problem remains to be solved.  The problem of study can be seen as a natural extension of  Bayraktar and Egami [10], and Avanzi, Shen and Wong [7]. In addition, one can see later that the method used in Bayraktar, Kyprianou and Yamazaki [11] cannot be applied to our model since their proof relies on certain characteristics of L\'evy process. In order to solve the optimal control problem in this paper, we shall first consider two sub-optimal problems in the next two sections.

 \vskip 0.2cm
\section{\normalsize Optimal dividend problem without capital injections}\label{without}
\setcounter{equation}{0}

In this section, we first consider the dividend problem
without capital injections. We shall show that the barrier strategy
solve the optimal dividend problem regardless of the jump distribution.

Let $\Xi_d=\{\xi_d=(L^{\xi_d}, G^{\xi_d}): (L^{\xi_d}, G^{\xi_d})\in
\Xi\; {\rm and}\; G^{\xi_d}\equiv 0\}$. The  associated controlled
process is denoted by
 $U^{\xi_d}=\{U_t^{\xi_d},t\ge 0\}$, where $U_t^{\xi_d}$ is the solution  to the stochastic
differential equation
\begin{equation}
\text{d}U^{\xi_d}_t=\text{d}P_t+ U^{\xi_d}_{t-}\text{d}R_t-\text{d}L_t^{\xi_d},\;\; t\ge 0.\nonumber
\end{equation}
and the value function is given by
\begin{equation}
V_{d}(x)=\sup_{ \xi_d\in\Xi_d}V(x; \xi_d)\equiv \sup_{
\xi_d\in\Xi_d}E_x\left(\alpha\int_{0-}^{\tau_{\xi_d}-} e^{-\delta
t}dL_t^{\xi_d}\right),\; x\ge 0, \label{without-eq1}
 \end{equation}
where $\tau_{\xi_d}=\inf\{t: U_t^{\xi_d}=0\}$ is the time of ruin under
the strategy $\xi_d$. We next identify the form of the value function
$V_d$  and the optimal strategy $\xi_d^*$ such that $V_d(x)=V(x;
\xi_d^*)$.

\noindent {\bf 3.1 HJB equation and  verification lemma}

For notational convenience, denote $v(x)=V(x; \xi_d^*)$. If $v$ is twice
continuously differentiable, then applying standard arguments from stochastic control theory (see Fleming and Soner [17]) or an approach similar to that in Azcue and Muler [9], we can show that the value function fulfils the dynamic programming principle
$$v(x)=\sup_{\xi_d\in\Xi}E_x\left(\int_0^{{\tau_{\xi_d}}\wedge T}e^{-\delta s}dL_s^{\xi_d}+e^{-\delta (\tau_{\xi_d}\wedge T)}v(U_{\tau_{\xi_d}\wedge T}^{\xi_d})\right),$$
for any stopping time $T$, and that the associated
Hamilton-Jacobi-Bellman (HJB) equation is
\begin{equation}
\max\{{\cal L} v(x)-\delta v(x),\; \alpha-v'(x)\}=0, \;
x>0,\label{without-eq2}
\end{equation}
 with $v(0)=0$, where ${\cal L}$ is the  the extended generator of $U$  defined
in (\ref{math-eq1}). The HJB equation (\ref{without-eq2}) can  also be obtained by the  heuristic argument of  Avanzi et al. [7].

\begin{lemma} (Verification Lemma) \
 Let   $v$ be a  solution to (3.2). Then, $v(x)\ge V(x; \xi_d)$ for any admissible strategy $\xi_d\in\Xi_d$,
 and thus  $v(x)\ge V_{d}(x).$
\end{lemma}
\noindent{\bf Proof.}\; For any admissible strategy $\xi_d\in\Xi_d$, put
$\Lambda=\{s: L^{\xi_d}_{s-}\neq L^{\xi_d}_s\}$.  Applying
Ito's formula  for semimartingale to $e^{-\delta t}v(U^{\xi_d}_{t})$
gives
\begin{eqnarray}
E_x[e^{-\delta (t\wedge \tau_{\xi_d}-)}v({U}^{\xi_d}_{t\wedge \tau_{\xi_d}-})]
&=&v(x)+ E_x\int_0^{t\wedge \tau_{\xi_d}-} e^{-\delta
s}({\cal L}-\delta)v({U}^{\xi_d}_{s-})ds \nonumber\\
&&+ E_x\sum_{s\in \Lambda,s\le t\wedge \tau_{\xi_d}-}e^{-\delta s}\left\{v({U}^{\xi_d}_{s})-v({U}^{\xi_d}_{s-})\right\} \nonumber \\
&&- E_x\int_{0-}^{t\wedge \tau_{\xi_d}-} e^{-\delta
s}v'({U}^{\xi_d}_{s-})d{L}^{\xi_d,c}_s,
\end{eqnarray}
where ${L}^{\xi_d,c}_s$ is the continuous part of ${L}^{\xi_d}_s$.
From (3.2), we see that $({\cal L}-\delta)v({U}^{\xi_d}_{s-})\le 0$ and $v'(x)\ge \alpha$.
Thus, for $s\in \Lambda,s\le t\wedge \tau_{\xi_d}$,
\begin{equation}
v({U}^{\xi_d}_{s})-v({U}^{\xi_d}_{s-})\le -\alpha ({L}^{\xi_d}_s-{L}^{\xi_d}_{s-}).
\end{equation}
It follows  from (3.3) and (3.4) that
\begin{equation}
E_x[e^{-\delta (t\wedge \tau_{\xi_d}-)}v({U}^{\xi_d}_{t\wedge \tau_{\xi_d}-})]\le
v(x)-\alpha  E_x \int_{0-}^{t\wedge \tau_{\xi_d}-} e^{-\delta s}d{L}^{\xi_d}_s.
\end{equation}
Letting $t\to\infty$ in (3.5) yields the result.   \hfill $\Box$

 \noindent {\bf 3.2 Construction of a candidate solution}

 It is assumed that dividends are paid according to the barrier
strategy $\xi_b$. Such a strategy has a level of  barrier  $b>0$.
When the surplus exceeds the barrier, the excess is paid out
immediately as dividends. Let $L_t^b$ be the total amount of
dividends up to time $t$. The controlled risk process when taking
into account of the dividend strategy  $\xi_b$ is $U^{b}=\{U_t^b,
t\ge 0\}$,
 where $U_t^b$ is  the solution to the following stochastic differential equation
\begin{equation}
\text{d}U^b_t=\text{d}P_t+ U^b_{t-}\text{d}R_t-\text{d}L_t^b,\;\; t\ge 0.\nonumber
\end{equation}
  Denote by $V_b(x)$ the expected discounted dividends
  function if the barrier strategy $\xi_b$ is applied, that
is,
\begin{equation}
V_b(x)=\alpha E_x \left( \int_{0-}^{T^x_b-}e^{-\delta t}dL^b_t\right),
\end{equation}
where $\delta>0$ is the force of interest and $T^x_b=\inf\{t\ge0: U^b_t=0\}.$

 The following result shows that $V_b(x)$ as a function of $x$
satisfies  an integro-differential equation with certain boundary
conditions.

\begin{lemma}\label{thrm3-1} For the risk process $U$ of   (2.3)  and the infinitesimal generator $\cal{L}$ of  (2.4),
 if $h_b(x)$ solves
$${\cal{L}} h_b(x)=\delta h_b(x), \quad  0< x <b,$$
and  $h_b(x) = h_b(b) + \alpha(x-b)$, for $x>b$, together with the boundary conditions
 $$h_b(0) = 0, \quad h_b'(b) = \alpha,$$
then $h_b(x)$ coincides with  $V_b(x)$ given by (3.6).
\end{lemma}
\noindent {\bf Proof.} \   Applying Ito's formula  for
semimartingale to $e^{-\delta t}h_b(U^b_{t-})$ gives
\begin{eqnarray}
e^{-\delta t}h_b({U}^{b}_{t-})&-&h_b({U}^{b}_0) = \int_{0-}^{t-}
e^{-\delta t}dN_s^{b}+\int_0^t e^{-\delta
s}({\cal L}-\delta)h_b({U}^{b}_{s-})ds \nonumber \\
&+ &\sum_{s< t}\text{\bf 1}_{\{\triangle
{L}_s>0\}}e^{-\delta s}\left\{h_b({U}^{b}_{s-}+\triangle
P_s-\triangle {L}_s)-h_b({U}^{b}_{s-}+\triangle
P_s)\right\} \nonumber \\
&-& \int_{0-}^{t-} e^{-\delta
s}h_b'({U}^{b}_{s-})d{L}_s^{c},\label{main-eq3}
\end{eqnarray}
where ${L}_s^{c}$ is the continuous part of ${L}_s$, and
\begin{eqnarray*}
N_t^{b}&=&\sum_{s\le t}\text{\bf 1}_{\{|\triangle
P_s|>0\}}\left\{h_b({U}^{b}_{s-}+\triangle P_s)-h_b({U}^{b}_{s-})
\right\}\\
 &&-\int_0^t\int_{0}^{\infty}\left\{h_b({U}^{b}_{s-}+y)
 -h_b({U}^{b}_{s-})\right\}\Pi(dy)ds\\
&&+\sigma \int_0^t h_b'({U}^{b}_{s-})dW_s.
\end{eqnarray*}
Note that $P(\triangle{L}_s>0, \triangle P_s<0)=0$ and that
${U}^{b}_{s-}+\triangle P_s\ge  {U}^{b}_{s-}+\triangle P_s-\triangle
{L}_s\ge b $ on $\{\triangle {L}_s>0, \triangle P_s>0\}$.
Consequently,
\begin{eqnarray*}
&\sum_{s< t}\text{\bf 1}_{\{\triangle {L}_s>0\}}e^{-\delta
s}\left\{h_b({U}^{b}_{s-}+\triangle P_s-\triangle
{L}_s)-h_b({U}^{b}_{s-}+\triangle P_s)\right\}\\
&= -\alpha\sum_{s<t}\text{\bf 1}_{\{\triangle {L}_s>0\}}e^{-\delta
s}\triangle {L}_s.
\end{eqnarray*}
Note that  $N_t^b$ is a local martingale, and
  $$\int_{0-}^{t-} e^{-\delta
s}h_b'({U}^{b}_{s-})d{L}_s^{c}= \int_{0-}^{t-}e^{-\delta
s}h_b'({U}^{b}_{s})d{L}_s^{c}=\alpha \int_{0-}^{t-} e^{-\delta
s}h_b'(b)d{L}_s^{c}.$$  Thus, for any appropriate localization
sequence of stopping times $\{t_n, n\ge 1\}$, we have
\begin{equation}
E_x (e^{-\delta (t_n\wedge T^{b})}h_b({U}^{b}_{t_n\wedge
T^{b}}))-E_x h_b({U}^{b}_0)= -\alpha E_x\int_{0-}^{t_n\wedge T^{b}-}
e^{-\delta s} d {L}_s.\label{main-eq4}
\end{equation}
Letting $n\to\infty$ in (\ref{main-eq4}) yields the result.   \hfill $\Box$

\begin{lemma} \label{without-4}
$V_{b}(x)$ is a concave increasing function on $(0, \infty)$.
\end{lemma}
\noindent{\bf Proof.} \  To prove the lemma, we use   arguments similar to those  in Kulenko and Schmidli [20]. Let $x>0, y>0$, and $l\in (0,1)$. Consider the strategies $L^x$ and $L^y$ for the initial capitals $x$ and $y$.
Define $L_t=l L^x_t+(1-l) L^y_t$. Then, $L_t=L_t^{lx+(1-l)y}.$ Since the processes $\{P_t, t\ge 0\}$ and  $\{R_t, t\ge 0\}$
 have no negative jumps, we have
$\tau_{L}=\tau_{L^x}\vee \tau_{L^y}$. It follows that
 \begin{eqnarray*}
 V_b(lx+(1-l)y)&=&\alpha E_x \left(\int_{0-}^{{\tau_L}-}e^{-\delta t}dL_t\right)\\
 &=& \alpha l E_x \left(\int_{0-}^{{\tau_L}-}e^{-\delta t}dL^x_t\right)+\alpha (1-l) E_x \left(\int_{0-}^{{\tau_L}-}e^{-\delta t}dL^y_t\right)\\
 &&\ge \alpha l E_x \left(\int_{0-}^{{\tau_{L^x}}-}e^{-\delta t}dL^x_t\right)
 + \alpha (1-l) E_x \left(\int_{0-}^{{\tau_{L^y}}-}e^{-\delta t}dL^y_t\right)\\
 &=&l V_b(x)+(1-l)V_b (y),
\end{eqnarray*}
and thus the concavity of $V_b$ follows. The increasingness of  $V_{b}(x)$ is trivial  \hfill $\Box$

\noindent {\bf 3.3 Verification of optimality}

 Define the barrier level by
$$b^*=\sup\{b\ge 0: V_{b}'(b-)=\alpha\}.$$
 We conjecture that the barrier strategy $\xi_{b^*}$ is optimal.
 \begin{proposition} \label{without-2} $b^*=0$ if and only if $\lambda\int_0^{\infty}y F(dy)\le p$.
\end{proposition}
\noindent {\bf Proof.} \ Here, we follow the approach of  Yao
et al. [23] to prove the proposition.    Suppose that
$b^*=0$. Then, the associated value function is
$V_d(x)=\alpha x$ which satisfies the HJB equation  (3.2).  As a result, we obtain $(\Gamma -\delta)V_d(x)\le 0$ which in turn gives $\lambda\int_0^{\infty}y F(dy)\le p$. On the other hand, suppose that
  $\lambda\int_0^{\infty}y F(dy)\le p$. Then,  $w(x)=\alpha x$ satisfies (3.2).
By Lemma 3.1, we get $w(x)\ge V_{d}(x)$. However,
$w(x)\le V_{d}(x)$ since $w(x)=\alpha x$ is the performance function associated with the strategy that $x$ is paid immediately as dividends. In this case, ruin occurs immediately.  Thus, $w(x)= V_{d}(x)$ and the optimal
barrier level $b^*=0$.  \hfill $\Box$

\begin{theorem} \label{without-5}\; If $\lambda\int_0^{\infty}y F(dy)>p$,  then the function $V_{b^*}$ defined in (3.6) satisfies
$$V_{b^*}(x)=V_d (x), \quad x\ge 0,$$
 and the optimal barrier strategy $\xi_{d}^{*}$  is the solution to
 \begin{eqnarray*} &&
\text{d}U_t^{\xi^*_d}=\text{d}P_t+  U_{t-}^{\xi^*_d}\text{d}R_t-\text{d} L_t^{\xi^*_d}, \qquad t\ge 0,
\end{eqnarray*}
with the conditions
\begin{eqnarray*}
U_t^{\xi^*_d}\le b^*, \quad G_t^{\xi_{d}^{*}}\equiv 0, \quad \int_0^{\infty}{\bf 1}_{\{U_s^{\xi^*_d}<b^*\}}dL_s^{\xi_d^*}=0.
\end{eqnarray*}
\end{theorem}
\noindent{\bf Proof.} \  Using the  method of Avanzi and Gerber [5], it can be shown that $V_{b^*}(x)$ is twice continuously
differentiable at $x=b^*$. Consequently,  $V_{b^*}\in
C^2(\Bbb{R}_+)$. Note that $({\cal L}-\delta)V_{b^*}(x)=0$ and $V_{b^*}'(x)\ge \alpha$ for $x\in [0,b^*)$ due to the
concavity of $V_{b^*}$ on $[0,b^*)$.  Since $V_{b^*}(x)=\alpha(x-b^*)+ V_{b^*}(b^*)$ for $x\ge b^*$, we have
\begin{eqnarray*}
({\cal L}-\delta)V_{b^*}(x)&=&-p\alpha+\alpha\int_0^{\infty}y F(dy)-\alpha (x-b^*)-\delta V_{b^*}(b^*)\\
&&<-p\alpha+\alpha\int_0^{\infty}y F(dy)-\delta V_{b^*}(b^*)\\
&=&\lim_{x\to b^*+}({\cal L}-\delta)V_{b^*}(x)=\lim_{x\to
b^*-}({\cal L}-\delta)V_{b^*}(x)=0,
\end{eqnarray*}
because of  the continuity of $V_{b^*}, V_{b^*}'$,  and
$V_{b^*}''$ at $x=b^*$. Thus, the function $V_{b^*}$ satisfies the HJB equation (3.2).
Then, it follows from Lemma 3.1 that $ V_{b^*}(x)\ge V_{d}(x)$. However,
$V_{b^*}(x)\le V_{d}(x)$ by definition, and hence $V_{b^*}(x)=V_{d}(x)$.    \hfill $\Box$

\noindent {\bf 3.4 Two closed-form solutions}

Owing to the complexity of the equation, the solution may not be available in explicit form in general. The following two examples show that one can derive closed-form solution in some special cases.

\noindent  {\bf Example 3.1.}\  Assume that $r=0$ and $\sigma_R=0$.  Then, $V_{b^*}(x)$ satisfies the following
  integro-differential equation
  \begin{equation}
  {\cal A}V_{b^*}(x)=\delta V_{b^*}(x), \quad 0<x<b^*,
  \end{equation}
and
 \begin{equation}
  V_{b^*}(x)=\alpha(x-b^*)+ V_{b^*}(b^*), \quad x>b^*,
\end{equation}
with the boundary conditions
  \begin{equation}
  V_{b^*}(0)=0, \quad   {V_{b^*}}'(x)|_{x=b^*}=\alpha,
   \end{equation}
  where
 \begin{equation}
  {\cal{A}} g(x)=\frac{1} {2}\sigma^2_{p} g''(x)-p g'(x)-\lambda g(x)
  +\lambda\int_{0}^{\infty}g(x+y)F(dy). \nonumber
  \end{equation}

Following the arguments of Laplace transform used in Yin, Wen and Zhao
[28], one can show that the solution to (3.9)-(3.11) is given by
$$ V_{b^*}(x)=-\alpha\overline{Z}^{(\delta)}(b^*-x)+\alpha\frac{E[X_1]}{\delta},$$
and
 $$b^*=(\overline{Z}^{(\delta)})^{-1}\left(\frac{E[X_1]}{\delta}\right),$$
  where
  $$Z^{(\delta)}(x)=1+\delta\int_0^x W^{(\delta)}(y)dy,\;
  \overline{Z}^{(\delta)}(x)=\int_0^x Z^{(\delta)}(y)dy, \ x\in \Bbb{R}.$$
 Here, $W^{(\delta)}$ is the so-called $\delta$-scale function defined in the way that $W^{(\delta)}(x) = 0$ for all $x < 0$ and that its Laplace transform on $[0,\infty)$ is given by
   \begin{equation}
   \int_0^{\infty}\text{e}^{-\theta
   x}W^{(\delta)}(x)dx=\frac{1}{\Psi(\theta)-\delta},\; \theta
   >\sup\{\theta\ge 0: \Psi(\theta)=\delta\},  \nonumber
   \end{equation}
 where
 $$ \Psi (\theta) =p\theta + \frac1 2\sigma^2_{p}\theta^2
+\lambda\int_{0}^{\infty}(e^{-\theta x}-1)F(dx). $$
For further details, the reader is referred to  Yin and Wen [26].  \hfill $\Box$

\noindent {\bf Example 3.2.}\ Let $\sigma_R=\sigma_p=0$. Assume that $X_i$ is exponentially distributed with parameter $\mu$.
 Then, by Theorem 3.1 and Lemma 3.2, it can be shown that $V_{b^*}(x)$ satisfies the following
  integro-differential equation
  \begin{equation}
  (rx-p)V_{b^*}'(x)+\lambda\mu\int_0^{\infty}V_{b^*}(x+z)e^{-\mu z}dz=(\lambda+\delta) V_{b^*}(x), \quad 0<x<b^*,
  \end{equation}
and
 \begin{equation}
  V_{b^*}(x)=\alpha(x-b^*)+ V_{b^*}(b^*), \quad x>b^*,
\end{equation}
with the boundary conditions
  \begin{equation}
  V_{b^*}(0)=0, \quad   {V_{b^*}}'(x)|_{x=b^*}=\alpha.
   \end{equation}

From equation (3.12), we find that
$$zg''(z)+\left(1-\frac{\lambda+\delta}{r}-z\right)g'(z)+\frac{\delta}{r}g(z)=0,$$
where
$$g(z)= V_{b^*}(x),\; z=\mu\left(x-\frac{p}{r}\right).$$
Note that this is Kummer's confluent hypergeometric equation with the solution given by
$$g(z)=C_1 M\left(-\frac{\delta}{r}, 1-\frac{\lambda+\delta}{r},z\right)+C_2 U\left(-\frac{\delta}{r}, 1-\frac{\lambda+\delta}{r},z\right),$$
where $C_1$ and $C_2$ are constants, and $M(a,b,x)$ is the standard confluent hypergeometric function with $U(a,b,x)$ being its second form; see, for example, Abramowitz and Stugen [1, pp. 504-505].  Then, it follows that
$$ V_{b^*}(x)=C_1 M\left(-\frac{\delta}{r}, 1-\frac{\lambda+\delta}{r}, \mu(x-\frac{p}{r})\right)
+C_2 U\left(-\frac{\delta}{r}, 1-\frac{\lambda+\delta}{r},\mu(x-\frac{p}{r})\right).$$
Using the boundary conditions (3.14) and the formulae
$$M'(a,b,z)=\frac{a}{b}M(a+1,b+1,z), \quad U'(a,b,z)=-aU(a+1,b+1,z),$$
we obtain the coefficients
$$C_1=\frac{\alpha U(-\frac{\delta}{r}, 1-\frac{\lambda+\delta}{r}, -\frac{\mu p}{r})}{\Delta(b^*)},$$
and
$$C_2=-\frac{\alpha M(-\frac{\delta}{r}, 1-\frac{\lambda+\delta}{r}, -\frac{\mu p}{r})}{\Delta(b^*)},$$
where
\begin{eqnarray*}
\Delta(b^*)&=&-\frac{\mu\delta}{r-\lambda-\delta}U\left(-\frac{\delta}{r},1-\frac{\lambda+\delta}{r}, -\frac{\mu
p}{r}\right)M\left(1-\frac{\delta}{r},
2-\frac{\lambda+\delta}{r},\mu(b^*-\frac{p}{r})\right)\\
&&+\frac{\mu\delta}{r}M\left(-\frac{\delta}{r},
1-\frac{\lambda+\delta}{r}, -\frac{\mu
p}{r}\right)U\left(1-\frac{\delta}{r},
2-\frac{\lambda+\delta}{r},\mu(b^*-\frac{p}{r})\right),
\end{eqnarray*}
and $b^*$ is the maximizer of term $1/\Delta(b)$ with
respect to $b$, i.e.,
$$b^*=argmax\frac{1}{\Delta(b)}.$$  \hfill $\Box$

 \vskip 0.2cm
\section{\normalsize Optimal dividend problem with capital injections}\label{with}
\setcounter{equation}{0}

In this section,  we consider the optimal dividend problem with
capital injections. The set of admissible strategies is given by
$$\Xi_c=\{\xi_c=(L^{\xi_c}, G^{\xi_c}): (L^{\xi_c}, G^{\xi_c})\in \Xi\; {\rm and}\; U_t^{\xi_c}\ge 0\}.$$
The controlled surplus process $U^{\xi_c}_t$ satisfies
\begin{equation}
\text{d}U^{\xi_c}_t=\text{d}P_t+ U^{\xi_c}_{t-}\text{d}R_t-\text{d}L_t^{\xi_c}+\text{d}G_t^{\xi_c}, \quad t\ge 0,\nonumber
\end{equation}
and the value function is defined as
\begin{equation}
V_{c}(x)=\sup_{ \xi_c\in\Xi_c}V(x; \xi_c)
\equiv \sup_{ \xi_c\in\Xi_c}E_x\left(\alpha\int_{0-}^{\infty}
e^{-\delta t}dL_t^{\xi_c}-\beta \int_{0-}^{\infty}\text{e}^{-\delta
t}dG_t^{ \xi_c}\right), \quad  x\ge 0.
\end{equation}
Since the controlled surplus process always stays positive, the company
will never go bankrupt. We shall identify the form of the value
function $V_c$  and the optimal strategy $\xi_c^*$ such that
$V_c(x)=V(x; \xi_c^*)$.

\noindent {\bf 4.1  HJB equation and verification lemma}

Applying  the  techniques used in Section 3, we get the HJB
equation and the verification Lemma.
\begin{equation}
\max\{{\cal L} w(x)-\delta w(x),\; \alpha-w'(x), w'(x)-\beta \}=0,
\quad x\ge 0.\label{with-eq2}
\end{equation}

\begin{lemma} (Verification Lemma) \
 Let   $w$ be a  solution to (4.2). Then, $w(x)\ge V(x; \xi_c)$ for any admissible strategy $\xi_c\in\Xi_c$,
 and thus  $w(x)\ge V_{c}(x).$
\end{lemma}

 \noindent {\bf 4.2 Construction of a candidate solution}

 We now construct a concave $C^2$ solution $H$ to the HJB equation (\ref{with-eq2}). Due to the effect of the discount factor, it is clear that the optimal strategy is the one that postpone capital injections as long as possible, i.e., we inject capital only when surplus become zero.
  Consider the barrier strategy with the upper barrier $B^*$ and the lower barrier $0$, and the strategy $\pi^*=(L^{\pi^*}, G^{\pi^*})$
  where $(U_t^{\pi^*}, L_t^{\pi^*, x}, G^{\pi^*, x})$ is a solution to the following system
 \begin{eqnarray}
 &&\text{d}U^{\pi^*}_t=\text{d}P_t+ U^{\pi^*}_{t-}\text{d}R_t-\text{d}L_t^{\pi^*}+\text{d}G_t^{\pi^*}, \\
 &&0\le U_t^{\pi^*}\le B^*, \ t\ge 0, \\
 &&L^{\pi^*,x}_t=\max(x-B^*,0)+\int_{0-}^{t-}1(U_s^{\pi^*}=B^*)d L_s^{\pi^*},\ t>0, \\
  &&G_t^{\pi^*,x}=\max\left(-\inf_{0\le s\le t}(P_s-L^{\pi^*}_s),0\right), \ t>0.
 \end{eqnarray}

 \begin{lemma}  For the problem of (4.3)-(4.6),  if $H(x)$ solves
$${\cal{L}} H(x)=\delta H(x), \quad  0< x <B^*,$$
with $H(x)=H(B^*) + \alpha(x-B^*)$ for $x>B^*$ and the boundary conditions
\begin{eqnarray*}
&&H'(0) = \beta, \quad H'(B^*) = \alpha,
\end{eqnarray*}
where  the  infinitesimal generator $\cal{L}$
is given by (2.4), then $H(x)$  is given by
\begin{equation}
H(x)=V(x; \pi^*)\equiv E_x\left(\alpha\int_{0-}^{\infty}
e^{-\delta t}dL_t^{\pi^*,x}-\beta \int_{0-}^{\infty}\text{e}^{-\delta
t}dG_t^{\pi^*,x}\right), \quad x\ge 0.
\end{equation}
\end{lemma}
\noindent{\bf Proof.} \ For the strategy $\pi^*$, define
$\Lambda=\{s: L^{\pi^*,x}_{s-}\neq L^{\pi^*,x}_s\}$.
Let  ${L}^{\pi^*,x,c}_t$ be the continuous part of ${L}^{\pi^*,x}_t$. Since the process is skip-free downward,
$G^{\pi^*,x}_{t}$ is continuous. In addition, we see from (4.6)  that $G^{\pi^*,x}_{t}\ge 0$ and that the support of the Stieltjes measure $dG^{\pi^*,x}_{t}$ is contained in the closure of the set $\{t: U_t^{\pi^*}=0\}$.
Applying Ito's formula  for semimartingale to $e^{-\delta t}H(U^{\pi^*}_{t})$
gives
\begin{eqnarray}
E_x[e^{-\delta t}H({U}^{\pi^*}_{t-})] &=&H(x)+ E_x\int_0^{t}
e^{-\delta
s}({\cal L}-\delta)H({U}^{\pi^*}_{s})ds \nonumber\\
&&+ E_x\sum_{s\in \Lambda,s\le t} e^{-\delta s}\left\{H(U^{\pi^*}_{s})
-H(U^{\pi^*}_{s-})\right\} \nonumber \\
&&- E_x\int_{0-}^{t-} e^{-\delta s}H'({U}^{\pi^*}_{s-})d{L}^{\pi^*,x,c}_s \nonumber\\
&&+E_x\int_{0-}^{t-} e^{-\delta s}H'({U}^{\pi^*}_{s-})d{G}^{\pi^*,x}_s.
\end{eqnarray}
Note that $({\cal L}-\delta)H({U}^{\pi^*}_{s})=0$, and that
$$E_x\sum_{s\in \Lambda,s\le t} e^{-\delta s}\left\{H(U^{\pi^*}_{s})
-H(U^{\pi^*}_{s-})\right\}=\alpha \sum_{s\le t} e^{-\delta s}({L}^{\pi^*,x}_s-{L}^{\pi^*,x}_{s-}),$$
$$E_x\int_{0-}^{t-} e^{-\delta s}H'({U}^{\pi^*}_{s-})d{L}^{\pi^*,x,c}_s=E_x\int_{0-}^{t-} e^{-\delta s}H'({U}^{\pi^*}_{s})d{L}^{\pi^*,x,c}_s=\alpha E_x\int_{0-}^{t-} e^{-\delta s}d{L}^{\pi^*,x,c}_s,$$
$$E_x\int_{0-}^{t-} e^{-\delta s}H'({U}^{\pi^*}_{s-})d{G}^{\pi^*,x}_s=E_x\int_{0-}^{t-} e^{-\delta s}H'({U}^{\pi^*}_{s})d{G}^{\pi^*,x}_s=\beta E_x\int_{0-}^{t-} e^{-\delta s}d{G}^{\pi^*,x}_s.$$
Then, it follows that
 \begin{equation}
E_x[e^{-\delta t}H({U}^{\pi^*}_{t-})] =H(x)-\alpha E_x\int_{0-}^{t-}
e^{-\delta s}d{L}^{\pi^*,x}_s+\beta E_x\int_{0-}^{t-} e^{-\delta
s}d{G}^{\pi^*,x}_s.
\end{equation}
Since $\lim_{t\to\infty} E_x[e^{-\delta t}H({U}^{\pi^*}_{t-})]\le
\lim_{t\to\infty} E_x[e^{-\delta t}H(B^*)]=0$, letting $t\to\infty$
in (4.9) and using the monotone convergence theorem yield
$$
H(x)=\alpha E_x\int_{0-}^{\infty} e^{-\delta s}d{L}^{\pi^*,x}_s-\beta E_x\int_{0-}^{\infty} e^{-\delta s}d{G}^{\pi^*,x}_s=V(x; \pi^*).
$$
\hfill $\Box$
\begin{lemma}
$V(x; \pi^*)$ is a concave increasing function on $(0, \infty)$.
\end{lemma}
\noindent{\bf Proof.} \ Similar to the proof of Lemma 3.3, we use the  arguments of   Kulenko and Schmidli [20]. Let $x>0$, $y>0$, and $l\in (0,1)$. Consider the strategies $(L^{\pi^*,x},G^{\pi^*,x})$ and $(L^{\pi^*,y},G^{\pi^*,y})$ for the initial capitals $x$ and $y$.
Define $L_t=l L^{\pi^*, x}_t+(1-l) L^{\pi^*, y}_t$ and $G_t=l G^{\pi^*,x}_t+(1-l) G^{\pi^*,y}_t$.  Then, $L_t=L_t^{\pi^*, lx+(1-l)y}$. So, we have
\begin{eqnarray*}
lx&+&(1-l)y
+\int_0^t{\cal E}(R)_{s-}^{-1}\text{d}P_s-\rho\sigma_p\sigma_R \int_0^t{\cal E}(R)_{s-}^{-1}\text{d}s\\
&&-\int_0^t  {\cal E}(R)_{s-}^{-1}(l dL^{\pi^*,x}_s+(1-l) dL^{\pi^*,y}_s)\\
&&+\int_0^t  {\cal E}(R)_{s-}^{-1}(l dG^{\pi^*,x}_s+(1-l) dG^{\pi^*,y}_s)\\
&=&  l\left\{x +\int_0^t{\cal E}(R)_{s-}^{-1}\text{d}P_s-\rho\sigma_p\sigma_R \int_0^t{\cal E}(R)_{s-}^{-1}\text{d}s\right. \\
&&\left.- \int_0^t  {\cal E}(R)_{s-}^{-1}dL^{\pi^*,x}_s+{\cal E}(R)_t\int_0^t  {\cal E}(R)_{s-}^{-1} dG^{\pi^*,x}_s\right\}\\
&&+ (1-l)\left\{y +\int_0^t{\cal E}(R)_{s-}^{-1}\text{d}P_s-\rho\sigma_p\sigma_R \int_0^t{\cal E}(R)_{s-}^{-1}\text{d}s\right.\\
&&- \left.\int_0^t  {\cal E}(R)_{s-}^{-1}dL^{\pi^*,y}_s+{\cal E}(R)_t\int_0^t  {\cal E}(R)_{s-}^{-1}dG^{\pi^*,y}_s\right\}\ge 0.
\end{eqnarray*}
This shows that the strategy $(L_t, G_t)$ is admissible and that $$G_t^{\pi^*,lx+(1-l)y}\le l G^{\pi^*,x}_t+(1-l) G^{\pi^*,y}_t.$$
It follows that
 \begin{eqnarray*}
 V(lx+(1-l)y,\pi^*)&=&E\left(\alpha \int_{0-}^{\infty}e^{-\delta t}dL_t^{\pi^*,lx+(1-l)y}- \beta\int_{0-}^{\infty}e^{-\delta t}dG_t^{\pi^*,lx+(1-l)y}\right)\\
 &&\ge l E \left( \alpha \int_{0-}^{\infty}e^{-\delta t}dL^{\pi^*,x}_t-\beta\int_{0-}^{\infty}e^{-\delta t}dG^{\pi^*,x}_t\right)\\
 &&+(1-l) E \left(\alpha \int_{0-}^{\infty}e^{-\delta t}dL^{\pi^*,y}_t-\beta\int_{0-}^{\infty}e^{-\delta t}dG^{\pi^*,y}_t\right)\\
 &=&l V(x,\pi^*)+(1-l)V(y,\pi^*),
\end{eqnarray*}
which implies the concavity of $V$. The   proof of increasingness of $V(x; \pi^*)$  is routine. \hfill $\Box$

\noindent {\bf 4.3 Verification of optimality}

Define the barrier level as
$$B^*=\sup\{B\ge 0: H'(B-)=\alpha\}.$$
 We conjecture that the barrier strategy $\pi^*$ is optimal.

\begin{theorem} \label{with-1}  The value function $H$ defined in (4.7)
satisfies
$$H(x)=V_c (x)=\sup_{\xi_c\in\Xi_c}V_{ \xi_c}(x),$$
 and the joint  strategy $\pi^*=(L^{\pi^*}, G^{\pi^*})$  is  optimal,
 where $(L^{\pi^*}, G^{\pi^*})$ is given by (4.5) and (4.6).
\end{theorem}

\noindent{\bf Proof.} \ Note that $({\cal L}-\delta)H(x)=0$ and $\alpha\le H'(x)\le \beta$ for $x\in [0,B^*)$ due to the
concavity of $H$ on $[0,B^*)$.  For $x\ge B^*$ and
$H(x)=\alpha(x-B^*)+ H(B^*)$, we have
\begin{eqnarray*}
({\cal L}-\delta)H(x)&=&-p\alpha+\alpha\int_1^{\infty}y\Pi(dy)-\alpha (x-B^*)-\delta H(B^*)\\
&&<-p\alpha+\alpha\int_1^{\infty}y\Pi(dy)-\delta H(B^*)\\
&=&\lim_{x\to b^*+}({\cal L}-\delta)H(x)=\lim_{x\to B^*-}({\cal
L}-\delta)H(x)=0.
\end{eqnarray*}
Due to the continuity of $H, H'$ and $H''$ at $x=B^*$.
Thus, the function $H$ satisfies the HJB equation (4.2).
By Lemma 4.1,  we get $H(x)\ge V_{c}(x)$. On the other hand,
$H(x)\le V_{c}(x)$. Thus,  $H(x)=V_{c}(x)$.  \hfill $\Box$

\noindent {\bf 4.4 Two closed-form solutions}

We now present two examples in which closed-form solution can be derived.

\noindent  {\bf Example 4.1.}\  Assume that $r=0$ and $\sigma_R=0$.  Then, $H(x)$ satisfies the following
  integro-differential equation
  \begin{equation}
  {\cal A}H(x)=\delta H(x), \; 0<x<B^*,
  \end{equation}
   and
 \begin{equation}
  H(x)=\alpha(x-B^*)+ H(B^*),\ x>B^*,
\end{equation}
 with the boundary conditions
  \begin{equation}
  H'(0)=\beta, \quad   H'(B^*)=\alpha,
   \end{equation}
  where
 \begin{equation}
  {\cal{A}} g(x)=\frac{1} {2}\sigma^2_{p} g''(x)-p g'(x)-\lambda g(x)
  +\lambda\int_{0}^{\infty}g(x+y)F(dy). \nonumber
  \end{equation}

Again, using the arguments of Laplace transform, one can show that the solution to (4.10) and (4.11) is given by
$$ H(x)=-\alpha\overline{Z}^{(\delta)}(B^*-x)+\alpha\frac{E[X_1]}{\delta},$$
and
 $$B^*=({Z}^{(\delta)})^{-1}\left(\frac{\beta}{\alpha}\right),$$
  where $Z^{(\delta)}(x)$ and  $\overline{Z}^{(\delta)}(x)$ are defined in Example 3.1. In the case of $\alpha=1$, these formulae were
   obtained in  Bayraktar,  Kyprianou and Yamazaki [11] by using the fluctuation theory of spectrally positive L\'evy processes.

\noindent {\bf Example 4.2.}\  Let $\sigma_R=\sigma_p=0$. Assume that $X_i$ is exponentially distributed with parameter $\mu$.
 Then, by Theorem 4.1 and Lemma 4.2, $H(x)$ satisfies the following
  integro-differential equation
  \begin{equation}
  (rx-p)H'(x)+\lambda\mu\int_0^{\infty}H(x+z)e^{-\mu z}dz=(\lambda+\delta) H(x), \quad 0<x<B^*,
  \end{equation}
  and
 \begin{equation}
  H(x)=\alpha(x-B^*)+ H(B^*), \quad x>B^*,
\end{equation}
 with the boundary conditions
  \begin{equation}
  H'(0)=\beta,\ \ \  H'(B^*)=\alpha.
   \end{equation}
  Repeating  the steps in Example 3.2, we obtain
$$H(x)=C_3 M\left(-\frac{\delta}{r}, 1-\frac{\lambda+\delta}{r}, \mu(x-\frac{p}{r})\right)
+C_4 U\left(-\frac{\delta}{r}, 1-\frac{\lambda+\delta}{r},\mu(x-\frac{p}{r})\right).$$
The constants $C_3$ and $C_4$ can be determined from the boundary conditions (4.15).
Using  the formulae
$$M'(a,b,z)=\frac{a}{b}M(a+1,b+1,z), \quad U'(a,b,z)=-aU(a+1,b+1,z),$$
we get
$$C_3=\frac{\beta \Delta_4-\alpha \Delta_2}{\Delta_1\Delta_4-\Delta_2\Delta_3},$$
and
$$C_4=\frac{\alpha \Delta_1-\beta \Delta_3}{\Delta_1\Delta_4-\Delta_2\Delta_3},$$
where
\begin{eqnarray*}
\Delta_1&=&-\frac{\mu\delta}{r-\lambda-\delta}M\left(1-\frac{\delta}{r},2-\frac{\lambda+\delta}{r},-\frac{\mu p}{r}\right),\\
\Delta_2&=&\frac{\mu\delta}{r}U\left(1-\frac{\delta}{r},
2-\frac{\lambda+\delta}{r},-\frac{\mu p}{r}\right),\\
\Delta_3&=&-\frac{\mu\delta}{r-\lambda-\delta}M\left(1-\frac{\delta}{r},2-\frac{\lambda+\delta}{r},\mu(B^*-\frac{p}{r})\right),\\
\Delta_3&=&\frac{\mu\delta}{r}U\left(1-\frac{\delta}{r},
2-\frac{\lambda+\delta}{r},\mu(B^*-\frac{p}{r})\right).
\end{eqnarray*}
Here, $B^*$ is the unique solution to the following equation with respect to $b$:
$$-\frac{\mu\delta}{r-\lambda-\delta} C_3 M\left(1-\frac{\delta}{r},
2-\frac{\lambda+\delta}{r},\mu(b-\frac{p}{r})\right)+\frac{\mu\delta}{r}U\left(1-\frac{\delta}{r},2-\frac{\lambda+\delta}{r},\mu(b-\frac{p}{r})\right)=\alpha.$$

 \vskip 0.2cm
\section{\normalsize Solution to the problem without constraints}\label{general}
\setcounter{equation}{0}

 We now consider the  control problem (2.6) without any restrictions on capital injections.  In this case, ruin can occur and the time of ruin for a control strategy $\xi$ is defined as
$$\tau_{\xi}=\inf\{t: U^{\xi}_t=0\},$$
because of the diffusion and the skip-free downward surplus process.
Then, it follows from (3.1), (4.1) and (2.5) that for all $x\ge 0$,
$V_{\xi}(x)\ge \max\{V_d(x), V_c(x)\}$. We shall determine $V_*$ and
the optimal strategy $\xi^*$ such that $V_*(x)=V(x; \xi^*)$.

\noindent {\bf 5.1 Verification lemma}

For the control problem without any restrictions on capital injections, we get the following
associated HJB equation:
\begin{equation}
\max\{{\cal L} v(x)-\delta v(x), \quad \alpha-v'(x), v'(x)-\beta \}=0,
\; x\ge 0,\label{general-eq1}
\end{equation}
with the boundary condition
\begin{equation}
\max\{-v(0), v'(0)-\beta\}=0.
\end{equation}

\begin{lemma} \label{general-1} (Verification Lemma) If $v$ satisfies the HJB equation (5.1) with the boundary condition (5.2), then $v(x)\ge V_{\xi}(x)$ for any admissible policy $\xi$.
\end{lemma}
\noindent{\bf Proof.} \ For any admissible strategy $\xi\in\Xi$, put
$\Lambda=\{s: L^{\xi}_{s-}\neq L^{\xi}_s\}$.  Applying
Ito's formula  for semimartingale to $e^{-\delta t}v(U^{\xi}_{t})$
gives
\begin{eqnarray}
E_x[e^{-\delta (t\wedge \tau_{\xi})}v({U}^{\xi}_{t\wedge
\tau_{\xi}-})] &=&v(x)+ E_x\int_0^{t\wedge \tau_{\xi}-} e^{-\delta
s}({\cal L}-\delta)v({U}^{\xi}_{s-})ds \nonumber\\
&&+ E_x\sum_{s\in \Lambda,s\le t\wedge \tau_{\xi}-}e^{-\delta s}\left\{v({U}^{\xi}_{s})-v({U}^{\xi}_{s-})\right\} \nonumber \\
&&- E_x\int_{0-}^{t\wedge \tau_{\xi}-} e^{-\delta
s}v'({U}^{\xi}_{s-})d{L}^{\xi,c}_s\nonumber \\
&&+E_x\int_{0-}^{t\wedge \tau_{\xi}-} e^{-\delta
s}v'({U}^{\xi}_{s-})d{G}^{\xi}_s,
\end{eqnarray}
where ${L}^{\xi,c}_s$ is the continuous part of ${L}^{\xi}_s$.
We see from (5.1)  that $({\cal L}-\delta)v({U}^{\xi_d}_{s-})\le 0$ and
$\alpha\le v'(x)\le \beta$. Thus,
\begin{equation}
E_x\int_{0-}^{t\wedge \tau_{\xi}-} e^{-\delta
s}v'({U}^{\xi}_{s-})d{G}^{\xi}_s\le \beta E_x\int_{0-}^{t\wedge \tau_{\xi}-} e^{-\delta
s}d{G}^{\xi}_s,
\end{equation}
and for $s\in \Lambda,s\le t\wedge \tau_{\xi}$,
\begin{equation}
v({U}^{\xi}_{s})-v({U}^{\xi}_{s-})\le -\alpha ({L}^{\xi}_s-{L}^{\xi}_{s-}).
\end{equation}
It follows  from (5.3) and (5.5) that
\begin{equation}
E_x[e^{-\delta (t\wedge \tau_{\xi})}v({U}^{\xi}_{t\wedge
\tau_{\xi}-})]\le v(x)-\alpha  E_x \int_{0-}^{t\wedge \tau_{\xi}-}
e^{-\delta s}d{L}^{\xi}_s +\beta  E_x \int_{0-}^{t\wedge
\tau_{\xi}-} e^{-\delta s}d{G}^{\xi}_s.
\end{equation}
Finally, by letting $t\to\infty$ in (5.6) and noting that (by Fatou's lemma)
$$
\liminf_{t\to\infty}E_x[e^{-\delta (t\wedge
\tau_{\xi})}v({U}^{\xi}_{t\wedge \tau_{\xi}-})]\ge
 E_x[\liminf_{t\to\infty} e^{-\delta (t\wedge \tau_{\xi})}v({U}^{\xi}_{t\wedge \tau_{\xi}})]\ge v(0)
 E_x[e^{-\delta \tau_{\xi})}]\ge 0,
$$
we prove the lemma. \hfill $\Box$

 \noindent {\bf 5.2 Construction of a candidate solution}

For any $x\ge 0$, we set our candidate strategy to be
 \begin{equation}
  \xi^*=\left\{
  \begin{array}{ll} \xi_{d}^{*},& {\rm if}\;  V'_{b^*}(0)\le \beta,\\
     \xi_{c}^{*}, &  {\rm if}\;  H(0)\ge 0,
     \end{array}
  \right.\label{general-eq2}
\end{equation}
and our candidate solution to be
\begin{equation}
  V_{\xi^*}(x)=\left\{
  \begin{array}{ll} V_d(x),&  {\rm if}\; V'_{b^*}(0)\le \beta,\\
     V_c(x), &  {\rm if}\;H(0)\ge 0,
  \end{array}
  \right.\label{general-eq3}
\end{equation}
where $V_d$ and $V_c$ are given by (3.1) and (4.1), respectively, and $V_{b^*}$ and $H$ are given by (3.6) and (4.7), respectively.

\noindent {\bf 5.3 Verification of optimality}
\begin{theorem} \label{general-1}
  The  value function $V_{\xi^*}$
defined in (5.8) satisfies
$$V_{\xi^*}(x)=V_*(x)=\sup_{\xi\in\Xi}V(x;\xi),$$
 and the joint  strategy $\xi^{*}$ defined in (5.7) is  optimal.
\end{theorem}
\noindent{\bf Proof.}\; If $V_{b^*}'(0)\le \beta$,
then $V_{b^*}$ satisfies the equation (5.1) with the  condition (5.2). Hence, $V_{b^*}(x)\ge V_*(x)$. On the other hand,
$V_{b^*}(x)=V(x; \xi_d^*)\le V_d(x)$. It follows that $V_{\xi^*}(x)=V_{b^*}(x)=V_d(x)$. The optimality of $\xi_d^*$
is verified by Theorem 3.1.  If $ H(0)\ge 0$,
then $H$ satisfies the HJB equation (4.1), so that $H(x)\le V_c(x)$. Since $H$ also satisfies the equation (5.1) with the  condition (5.2),  $H(x)\ge V(x; \xi_c^*)\ge V_c(x)$. Hence, we have $V_{\xi^*}(x)=V(x; \xi_c^*)=V_c(x)$. The optimality of $\xi_c^*$ is verified by Theorem 4.1.
\hfill $\Box$

\vskip0.3cm

\noindent{\bf Acknowledgements}

The authors would like to thank two anonymous referees and the editor for their helpful comments on the previous version of the paper.
The research of Chuancun Yin was supported by the National Natural Science Foundation of China (No. 11171179) and the Research Fund for the Doctoral Program of Higher Education of China (No.  20133705110002). The research of Kam C. Yuen was supported by a grant from the Research Grants Council of the Hong Kong Special Administrative Region, China (Project No. HKU 7057/13P).

\end{document}